# Enhanced Kinetic Impactor for Deflecting Large Potentially Hazardous Asteroids via Maneuvering Space Rocks

**Mingtao Li**[1,2]*, **Yirui Wang**[1,2], **Youliang Wang**[1,2], **Binghong Zhou**[1,2], **Wei Zheng**[1]

[1]National Space Science Center, Chinese Academy of Sciences

[2]University of Chinese Academy of Sciences

*Correspondence Author. E-mail: limingtao@nssc.ac.cn

## Abstract

Asteroid impacts pose a major threat to all life on Earth. The age of the dinosaurs was abruptly ended by a 10-km-diameter asteroid. Currently, a nuclear device is the only means of deflecting large Potentially Hazardous Asteroids (PHAs) away from an Earth-impacting trajectory. The Enhanced Kinetic Impactor (EKI) concept is proposed to deflect large PHAs via maneuvering space rocks. First, an unmanned spacecraft is launched to rendezvous with an intermediate Near-Earth Asteroid (NEA). Then, more than one hundred tons of rocks are collected from the NEA as the EKI. The NEA can also be captured as the EKI if the NEA is very small. Finally, the EKI is maneuvered to impact the PHA at a high speed, resulting in a significant deflection of the PHA. For example, to deflect Apophis, as much as 200 t of rocks could be collected from a NEA as the EKI based on existing engineering capabilities. The EKI can produce a velocity increment ($\Delta v$) of 39.81 mm/s in Apophis, thereby increasing the minimum geocentric distance during the close encounter in 2029 by 1,866.93 km. This mission can be completed in 3.96 years with a propellant cost of 2.98 t. Compared with a classic kinetic impactor, the deflection distance can be increased one order of magnitude. The EKI concept breaks through the limitation of the ground-based launch capability, which can significantly increase the mass of the impactor. We anticipate that our research will be a starting point for efficient planetary defense against large PHAs.

## Introduction

The vast majority of known asteroids orbit within the main asteroid belt located between the orbits of Mars and Jupiter, but some asteroids pass in proximity to Earth, thereby threatening all life on Earth. Several serious impact events (i.e. Chixulub event[1], Tunguska event[2] and Chelyabinsk event[3]) have aroused people's attention to the research of planetary defense. The 2010 report[4] of the American Academy of Sciences summarizes the general range of different deflection strategies with respect to mission time and target size. Passive methods such as ground-based civil defense are the most effective methods to defend against small impact events. For potentially hazardous asteroids (PHAs) with short warning times or large masses, nuclear explosion is the only feasible option[5]. However, nuclear explosions may cause controversy. A gravitational tractor is insensitive to the structure, surface properties and rotation state of the asteroid[6], however, the main caveat is the requirement for the spacecraft propulsion system to operate reliably for perhaps a decade or more[4]. The ARM (Asteroid Redirect Mission) proposed by NASA in 2013 studied the feasibility of using a tractor to maneuver small asteroids[7-9], but this strategy does not apply to large asteroids.

Despite of fragmentation risks, a kinetic impactor remains a promising strategy for asteroid deflection[10]. In 2005, the Deep Impact mission released an impactor weighing 372 kg to collide with comet Tempel 1 at a velocity of 10.2 km/s[11]. This impact generated a 0.0001 mm/s velocity change in the comets orbital velocity and decreased its perihelion distance by 10 meters[12]. DART (Double Asteroid Redirection Test) is currently planned to be implemented in October 2022, with a 555 kg spacecraft hitting Didymos' moon at a relative velocity of 6.65 km/s, generating a velocity change of 0.8-2 mm/s (depending on the $\beta$, which measures an additional momentum transferred to the target by the thrust in the opposing direction of crater ejecta that escapes the gravitational attraction of the target body[13]). The DART mission will be the first demonstration of a kinetic impactor and will provide useful data for the technique[10]. Many studies have been performed to improve the deflection efficiency of a kinetic impactor, such as using retrograde orbit[14,15] or H-reversal orbit[16-18] to improve the impact velocity. However, due to the limited weight of the artificial impactor, using a spacecraft to hit a large asteroid is similar to using an egg to hit a rock. Even at a higher impact velocity, the improvement of the deflection efficiency for large asteroids is limited.

The Enhanced Kinetic Impactor (EKI) concept is proposed to deflect large PHAs via maneuvering space rocks. The EKI concept is described in four key stages as follows, and its schematic diagram is shown in **Figure 1**.

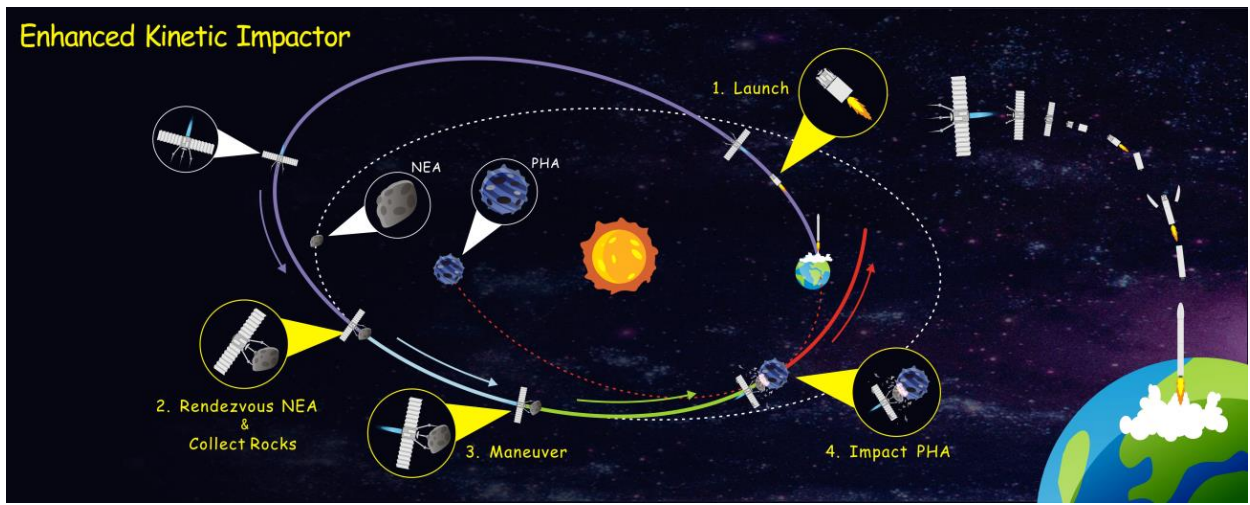

**Figure 1.** Schematic diagram of the Enhanced Kinetic Impactor (EKI). Four key actions are partially magnified in yellow circles: Launch, Rendezvous with a NEA & Collect Rocks, Maneuver, and Impact PHA. The details of the NEA, PHA and spacecraft are magnified in gray circles. The gray and red dotted lines indicate the original orbits of the NEA and PHA, respectively. Solid lines of different colors represent transfer orbits at different stages.

1）Launch. An existing heavy launch vehicle, e.g., Long March 5, is used to launch an unmanned spacecraft from Earth.

2）Rendezvous with a NEA & Collect Rocks. This stage refers to the ARM concept[9]. The unmanned spacecraft is used to rendezvous with an intermediate NEA. More than one hundred tons of rocks are collected from the NEA as the EKI, an entire NEA can also be captured as the EKI if the NEA is very small. In theory, the number of asteroids in 10 m diameter is as high as 100 million[19], which provides ample candidates to be captured as EKIs. However, due to the limited observation capability, the number of observed NEAs less than 10 m in diameter is approximately 1000[20]. Based on the collected data on asteroids, such as Itokawa and Ryugu, large asteroid surfaces feature an abundance of scattered rocks (boulders)[21,22]. Therefore, even if it is impossible to capture the entire NEA, we can collect rocks from the NEA as the EKI. For more technical details, such as how to capture a small asteroid and collect rocks, will be described in the discussion section.

3）Maneuver. After the rocks are assembled into the EKI, the electric propulsion system begins to maneuver the EKI away from the original orbit toward the PHA. During this process, some in-situ detections or experiments can be performed. In the future, the concept of EKI may also introduce the possibility of processing asteroid material to extract propellants.

4）Impact PHA. The EKI is maneuvered to impact the PHA at a high speed. After impact, the PHA will be farther away from Earth during the close encounter. In this paper, the deflection effect is evaluated by the change in the minimum geocentric distance, which is called the deflection distance.

By using space rocks as the impactor, the EKI concept has the following main advantages: 1) breaking through the limitation of the ground-based launch constraints, thereby increasing the mass of the impactor significantly with the current technology; 2) combining the areas of science, planetary defense and exploration, which can generate more scientific returns compared with classic kinetic impactor.

## Results

Apophis is considered to be one of the most hazardous PHAs. Its minimum geocentric distance will be 0.00025 AU in 2029[23]. Although the chance of an impact is currently predicted to be quite low, the case of Apophis is interesting to test defense techniques. To demonstrate the power of the EKI scheme, this paper takes Apophis as an example. We designed a deflection mission and analyzed the deflection effect for Apophis. The results of the design are shown in **Table 1**. The total flight time is 3.96 years.

According to the launch capability of Long March 5, the spacecraft has a total mass of 8 t with a propellant mass of 4 t. The spacecraft is equipped with a solar electric propulsion (SEP) system, similar to the ARM. The specific impulse of the SEP is 3,000 s, and the maximum thrust is 1.5 N. First, the spacecraft escapes from the Earth's sphere of influence (SOI) on Oct. 7, 2021. After 810 days, it rendezvouses with asteroid 2017 HF on Dec. 26, 2023. Then, the spacecraft works around the asteroid 2017 HF and collects 200 t of rocks as the EKI. Finally, the 200 t of rock are maneuvered to intercept Apophis starting on Sep. 8, 2024. After 380 days, the EKI composed of 200 t of rocks impacts Apophis on Sep. 23, 2025 with a relative velocity of 11.84 km/s. Assuming Apophis weighs 6.1x10$^{10}$ kg and $\beta = 1$, the $\Delta v$ of Apophis caused by the deflection is 39.81 mm/s. Apophis' minimum geocentric distance during the close encounter in 2029 increases from 0.000252 AU to 0.000265 AU, an increase of 1,866.93 km. The mission costs 2.98 t of propellant for transfer maneuvers, and has a 1.02 t redundancy.

**Table 1.** Results of mission design

| | |
|---|---|
| Intermediate NEA | 2017 HF |
| Mass of Spacecraft (Dry mass + Propellant mass) | 8 t (4 t + 4 t) |
| Specific Impulse of SEP | 3,000 s |
| Max Thrust of SEP | 1.5 N |
| Earth Escape Date | Oct. 7, 2021 |
| $C_3$ | 0 km$^2$/s$^2$ |
| "Earth-2017 HF" Flight time | 810 days |
| "Earth-2017 HF" Propellant Cost | 2.01 t |
| Asteroid Arrival Date | Dec. 26, 2023 |
| Rendezvous & Assembling Time | 257 days |
| Asteroid Escape Date | Sep. 8, 2024 |
| "2017 HF-Apophis" Flight Time | 380 days |
| "2017 HF-Apophis" Propellant Cost | 0.97 t |
| Impact Date | Sep. 23, 2025 |
| Mass of the Impactor (Rocks + Spacecraft) | 205.02 t (including 200 t of rocks) |
| Impact Relative Velocity | 11.84 km/s |
| $\Delta v$ of Apophis | 39.81 mm/s |
| $\Delta$ Perigee of Apophis in 2029 | +1,866.93 km |

The Standard Dynamical Model (SDM)[24] is used for propagation, which includes n-body relativistic gravitational forces caused by the Sun, planets, Moon, Ceres, Pallas, and Vesta. The lunar and planetary ephemerides are based on JPL DE430[25]. The asteroids ephemerides are downloaded from JPL Horizons On-Line Ephemeris System[26]. **Figure 2** shows a comparison of the deflected orbit (after impact) and the original orbit (before impact), where the ordinate indicates the distance between Apophis and Earth.

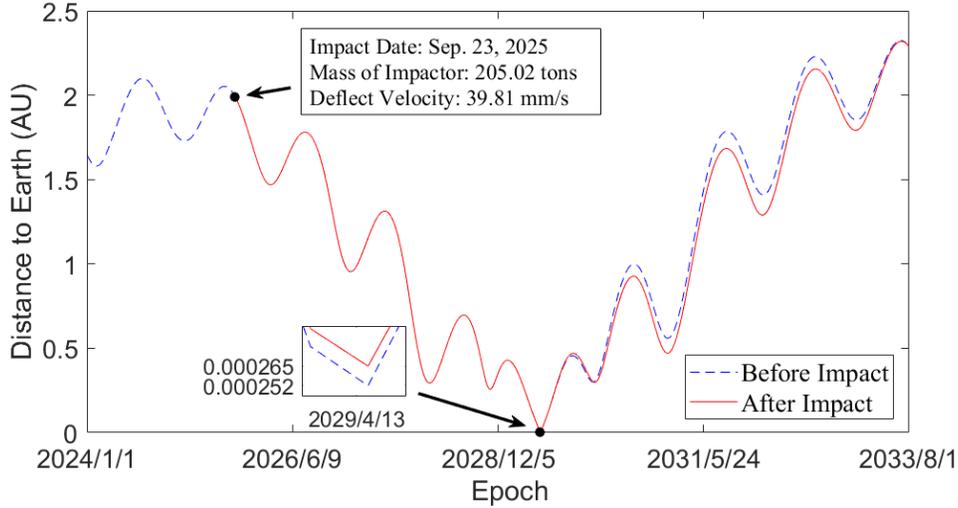

**Figure 2.** Distance between Apophis and Earth. The blue dotted line indicates the geocentric distance before Apophis is impacted. The red solid line indicates the geocentric distance after Apophis is impacted. The black dot on the left indicates the point at which Apophis is impacted by the EKI. The black dot on the right shows the details of the change in Apophis' minimum geocentric distance

## Discussion

**Efficiency of the Enhanced Kinetic Impactor**

In order to demonstrate the power of the EKI concept, we added a comparative simulation: optimizing the deflection efficiency of using a launch vehicle to send a spacecraft directly into the impact orbit to Apophis (also called Classic Kinetic Impactor, CKI). The deflection effects of CKI and EKI are compared.

A CKI mission is designed under the same initial conditions of the launch vehicle (Long March 5, CZ-5) and the warning time (10 years). We take the deflection distance (Δ Perigee of Apophis in 2029) as the objective function, and use Genetic Algorithm to optimize the launch date (Jan. 1, 2020 ~ Jan. 1, 2022) and transfer time (200~2000 days) for the CKI mission. The details for the CKI mission design method are described in the method section. **Table 2.** compares the mission design results of the CKI and the EKI.

**Table 2.** Mission design results of the two types of the kinetic impactor

| Type | Classic Kinetic Impactor | Enhanced Kinetic Impactor |
|---|---|---|
| Launch Vehicle | CZ-5 | CZ-5 |
| C3 | 23.72 km$^2$/s$^2$ | 0 km$^2$/s$^2$ |
| Mass of impactor | 3.56 tons | 205.02 tons |
| Launch date | May. 2, 2020 | Oct. 7, 2021 |
| Transfer time | 670.73 days | 1447 days |
| Impact date | Mar. 4, 2022 | Sep. 23, 2025 |
| Δ$v$ of Apophis | 0.38mm/s | 39.81 mm/s |
| Δ Perigee of Apophis in 2029 | +176.22 km | +1,866.93 km |

As we can see from **Table 2**, with the same launch vehicle and same warning time, the EKI can increase the $\Delta v$ of Apophis from 0.38 mm/s to 39.81 mm/s, the deflection distance from 176.22 km to 1,866.93 km. Compared with a CKI, the deflection distance can be increased one order of magnitude. The EKI breaks through the limitation of the ground-based launch constraints, which means that the mass of the impactor can be increased from several tons to more than one hundred tons. If a future heavy launch vehicle is considered, a more powerful impactor will be achieved.

**Operational Feasibility of the Enhanced Kinetic Impactor**

This paper focuses on the feasibility of the EKI from the perspective of mission analysis. For the technical issues, both two options of capturing a small asteroid and collecting rocks build on the heritage and lessons learned from the design of the ARM. A brief review of these technical issues is given to illustrate the operational feasibility of the EKI.

a) Rendezvous. This stage's technical issues refer to several real asteroids/comets missions, such as NEAR-Shoemaker mission to the Eros[27,28], the Hayabusa1/2 mission to Itokawa/Ryugu[29-32], the OSIRIS-Rex mission to Bennu[33,34], and the Rosetta mission to Churyumov–Gerasimenko[35].

b) Capture a small asteroid. As the design results of ARM-Option-A, after rendezvousing with the NEA, the spacecraft will observe it to determine the spin state, then the spacecraft will approach in line with the pole and spins up to match the rotating rate of the NEA. The spacecraft will move in slowly to enclose the target NEA in the canister. The floating drawstring motor would pull the bag around the NEA, securing it tightly to the spacecraft. At last, the spacecraft will use the hydrazine reaction control system (RCS) to de-spin the asteroid and spacecraft. The RCS concept is a single fault tolerant, hypergolic bipropellant subsystem using monomethyl hydrazine (MMH) and nitrogen tetroxide (NTO) with a gaseous nitrogen pressurization system. It includes four pods of four thrusters. Each thruster has a nominal thrust of 200 N and a specific impulse of 287 s. Assuming the target NEA is rotating at 1 RPM (Revolutions Per Minute) about its major axis, and has a cylindrical shape with a mass of 1,100 t. By using the RCS, the resulting time for de-spin will be ~33 minutes assuming continuous firing, and approximately 306 kg of propellant would be required[36].

c) Collect rocks. As the design results of ARM-Option-B, the design includes 2 capture arms and 3 contact arms. The 2 capture arms are used to fix the spacecraft to the rock, and the 3 contact arms are designed to absorb the momentum of the spacecraft using electrically driven linear actuators[37]. The actuators are selected from the Mars Exploration Rover (MER), and this landing method can prevent regolith and dust from being disturbed and settling on the solar arrays, optics, and other sensitive equipment[38]. After landing, 2 capture arms with microspine grippers with anchoring drills will then use hundreds of fishhook-like spines to opportunistically grab the surface features of the asteroid[39]. Microspines can attach to both convex and concave asperities, such as pits, protrusions, and sloping rock faces[40]. The Jet Propulsion Laboratory (JPL) conducted a large number of experiments on microspine grippers and studied their feasibility, as described in detail in the literature[39]. In terms of the capture capability, this capture system is scalable to larger boulders, with a boulder on the order of 10 meters in size required to provide approximately 1,000 metric tons of in situ mass[41].

d) Maneuvering. As the simulation results of Keck Institute for Space Studies (KISS), it was found that using a 40 kW SEP system operating four 3,000 s specific impulse Hall thrusters, an 1800 t asteroid could be returned to cis-lunar space within a ten-year mission span[36]. This shows the feasibility of maneuvering hundreds of tons of space rocks.

In summary, the processes of capturing and maneuvering space rocks, which have been fully studied by the ARM, are technically feasible. Our research in this essay mainly focuses on what we can do with the captured rocks based on the existing technology. For example, Mazanek, D. D. et al.[41] proposed using captured rocks to form an enhanced gravity tractor (EGT), which uses the collected rocks to augment the mass of a gravity tractor. In this paper, we propose the Enhanced Kinetic Impactor (EKI) concept, which uses the collected rocks to impact PHAs.

## Conclusions

This paper proposes a novel concept of deflecting large asteroid via maneuvering space rocks, which is called Enhanced Kinetic Impactor (EKI). A mission for deflecting Apophis is designed to demonstrate the feasibility and power of the EKI. The concept of EKI breaks through the limitation of the ground-based launch capability, which means that the mass of the impactor can be increased from several tons to more than one hundred tons. At the same time, the EKI mission can generate more scientific returns by combining the areas of science, planetary defense and exploration. The EKI concept can be used as an alternative method to defend large asteroids in the future.

## Methods

**Target Selection**
The first step of mission design is to determine an intermediate NEA to rendezvous with. The intermediate NEA has to have an orbit that allows a low $\Delta v$ for both trajectory legs. Due to the huge mass of the EKI, the preliminary selection is mainly based on the maneuver $\Delta v$ of the EKI after the rendezvous of the spacecraft and the NEA. The intermediate NEAs have been selected from JPL Small-Body Database, and the corresponding ephemerides[26] are used. First, we selected all the NEAs that satisfied $i < 10^\circ$ and $e < 0.3$. Then, by solving the Lambert problem, the optimal $\Delta v$ for transferring intermediate NEAs to Apophis was calculated. Finally, the NEAs with less $\Delta v$ were selected as the candidate intermediate NEAs. By using the SDM for propagation, corresponding deflection distance ($\Delta$ Perigee of Apophis in 2029) can be calculated (200 t of rocks are collected). **Table 3** shows seven candidate intermediate NEAs with $\Delta v$ values of less than about 0.1 km/s.

Table 3. Candidate intermediate NEA selection results

| Object | Maneuver date | $\Delta v$ (km/s) | Transfer time (days) | $\Delta$ Perigee of Apophis in 2029 (km) |
|---|---|---|---|---|
| 2014 HL198 | May. 11, 2025 | 0.022723210 | 675.73 | 1,110.58 |
| 2017 DR35 | May. 29, 2023 | 0.022911499 | 893.60 | 1115.73 |
| 2017 HF | Sep. 12, 2024 | 0.034142525 | 377.16 | 1,729.02 |
| 2006 DM | Nov. 6, 2025 | 0.065718796 | 889.90 | 936.61 |
| 1994 EU | Apr. 21, 2023 | 0.091977345 | 541.48 | 3,709.68 |
| 2018 BQ | Apr. 21, 2026 | 0.096481530 | 401.52 | 1,687.65 |
| 2000 AF205 | Oct. 8, 2023 | 0.122164737 | 823.77 | 1,335.69 |

The meanings of the items in the table are as follows. Applying $\Delta v$ to the object on the maneuver date is the optimal pulse transfer strategy, and the intermediate NEA will impact Apophis after the transfer time. The initial state is the position of the intermediate NEA, and the end state is the position of Apophis.

Comprehensively considering the $\Delta v$ and the deflection distance, we choose asteroid 2017 HF as the intermediate NEA, which is used to demonstrate the feasibility of the EKI. We plan to collect 200 t of rocks as the EKI from the asteroid 2017 HF and use the electric propulsion system to maneuver the space rocks. The collected rocks are assumed to have a spherical shape. The diameter of C-type ($\rho$ = 1.38 g/cm$^2$) rocks is 6.5 m, and the diameter of S-type ($\rho$ = 2.71 g/cm$^2$) rocks is 5.2 m.

**Low-Thrust Trajectory Design**

Since a solar electric propulsion (SEP) system is selected, this section gives a brief introduction of the design process of low-thrust trajectories. The whole transfer trajectories are divided into three arcs: "Earth-2017 HF", "Assemble Enhanced Kinetic Impactor (EKI)" and "2017 HF-Apophis". Each arc is described by the two-body problem. The equation of motion in the Cartesian coordinate system can be described as

$$\dot{\boldsymbol{r}} = \boldsymbol{v} \quad \dot{\boldsymbol{v}} = -\frac{\mu}{r^3}\boldsymbol{r} + \frac{T}{m}\boldsymbol{u} \quad \dot{m} = -\frac{T_{max}}{g_0 I_{sp}} \quad (1)$$

Where $\boldsymbol{r}$ is the position and $\boldsymbol{v}$ is the velocity, $\mu$ is the gravitational constant of center body, $m$ is the total mass of spacecraft, $T_{max}$ is the maximal thrust of the electric propulsion system, $g_0$ is the gravitational acceleration of Earth, $I_{sp}$ is the specific impulse.

An indirect method based on the theory of optimal control is applied to design the low-thrust transfer trajectories. To improve the convergence efficiency of the fuel-optimal problem of low-thrust trajectory, the homotopy technique[42] is used, which is starting from an easier problem (i.e. energy-optimal problem) and introducing a homotopic coefficient $\varepsilon$. The optimal problem can be described as

$$J = \frac{T_{max}}{g_0 I_{sp}} \int_{t_0}^{t_f} [u - \varepsilon(1-u)u] dt \quad (2)$$

$$\text{s.t. } \boldsymbol{r}(t_f) = \boldsymbol{r}_f, \quad \boldsymbol{v}(t_f) = \boldsymbol{v}_f$$

Where $0 \leq u \leq 1$, $0 \leq \varepsilon \leq 1$. $\varepsilon = 1$ indicates the energy-optimal problem, $\varepsilon = 0$ indicates the fuel-optimal problem. The optimal control $u^*$ can be calculated based on Pontryagin's maximum principle (PMP).

$$u^* = \begin{cases} 1, & \rho < -\varepsilon \\ \frac{1}{2} - \frac{\rho}{2\varepsilon}, & -\varepsilon < \rho < \varepsilon \\ 0, & \rho > \varepsilon \end{cases} \quad (3)$$

Where $\rho$ indicates the switch function

$$\rho = 1 - p_m - \frac{g_0 I_{sp}}{m} \|\boldsymbol{p}_v\| \quad (4)$$

Since the "Earth-2017 HF" arc has no constraints for terminal propellant mass, and the "2017 HF-Apophis" arc has no constraints for the terminal propellant mass and velocity. According to the transversal condition, this optimal control problem can be finally transformed into the following boundary value problem.

$$\Phi_{Rendezvous}(z) = [\boldsymbol{r}(t_f) - \boldsymbol{r}_f \quad \boldsymbol{v}(t_f) - \boldsymbol{v}_f \quad \lambda_m(t_f)] = 0 \quad (5)$$

$$\Phi_{Impact}(z) = [\boldsymbol{r}(t_f) - \boldsymbol{r}_f \quad \boldsymbol{\lambda}_v(t_f) \quad \lambda_m(t_f)] = 0 \quad (6)$$

The optimal solution (maneuver date and transfer time) of an impulsive trajectory is utilized to give the initial guess of the maneuver date and transfer time of low-thrust trajectory. The impact is performed at 2017 HF&Apophis Closest-Approach date (Sep. 23, 2025). "2017 HF-Apophis" arc costs 380 days, "Earth-2017 HF" arc costs 810 days, and "Assemble Enhanced Kinetic Impactor" arc costs 257 days. The trajectories are given in **Figure 3**. With the initial propellant mass of 4 t, the propellant cost of the whole transfer trajectory has been shown in **Figure 4**. The total flight time in this case would be 3.96 years.

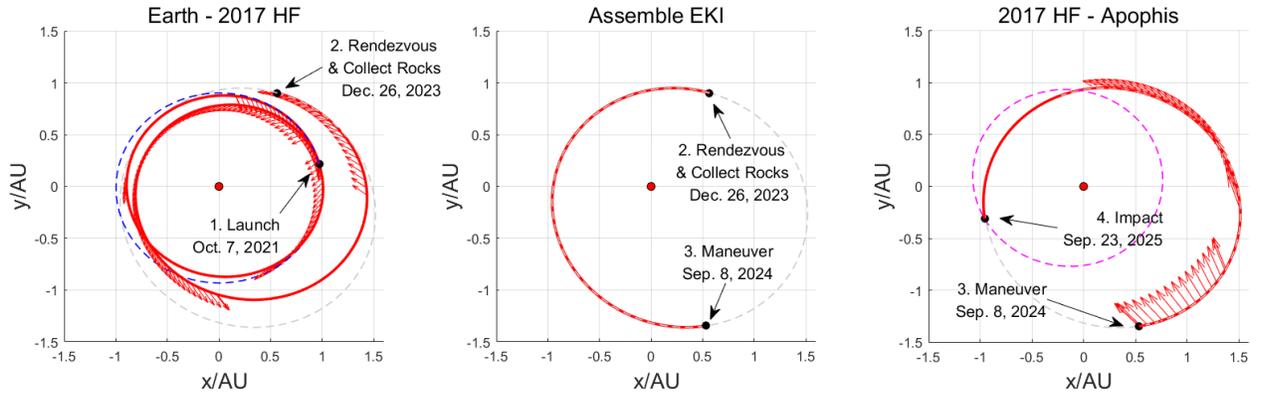

**Figure 3.** Fuel-optimal low-thrust transfer trajectories. The red dot indicates Sun. Blue, gray and magenta dotted lines indicate the orbits of Earth, 2017 HF, and Apophis. Red solid lines indicate low-thrust trajectories. The four key actions mentioned in **Figure 1** are noted in the figure.

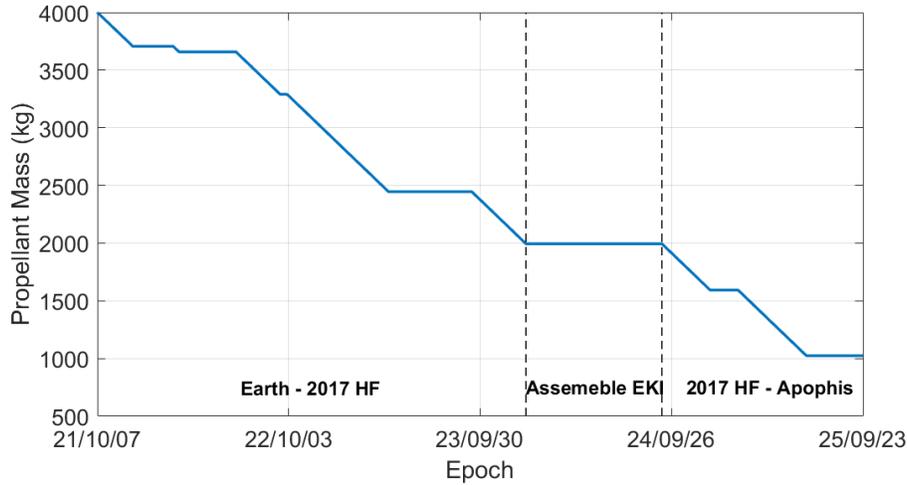

**Figure 4**. Propellant mass as a function of epoch. The two dotted lines divides the figure into three parts, which allows to visualize the propellant cost of the three arcs. The remaining propellant (1.02 t) can be used for other operations (de-spin et. al) and the mass of EKI.

**Deflection Effect**

Assuming that the impact process is a complete inelastic collision with two spheres, according to the law of conservation of momentum, the velocity increment of PHA caused by the impact is

$$\Delta \boldsymbol{v}_{PHA} = \beta \frac{m_{EKI}}{m_{EKI}+m_{PHA}} (\boldsymbol{v}_{EKI} - \boldsymbol{v}_{PHA}) \qquad (7)$$

Where $m_{EKI}$ and $\boldsymbol{v}_{EKI}$ indicate EKI's mass and velocity, $m_{PHA}$ and $\boldsymbol{v}_{PHA}$ indicates the PHA (Apophis) mass and velocity. $\beta$ indicates the effect of ejecta, which plays an important role of determining the final linear momentum of the asteroid. $\beta = 1$ indicates no ejecta contribution to the transferred momentum, and $\beta > 1$ indicates the incident momentum from the impactor is simply transferred to the Apophis has been enhanced by ejecta.

This paper assumes that the incident momentum from the impactor is simply transferred to the Apophis without enhancement ($\beta = 1$). The total mass of the EKI and the remaining propellant is 205.02 t, and the impact relative velocity of the EKI is (-9.16648, -7.45755 0.79576) km/s. The mass of Apophis is expected to be $6.1 \times 10^7$ t[20]. As a result, the $\Delta \boldsymbol{v}$ of Apophis caused by the impact is (-30.80895, -25.06515, 2.67458) mm/s, and the magnitude is 39.81 mm/s. The orbit elements of Apophis before and after the impact are shown in **Table 4**.

**Table 4**. Orbit Elements of Apophis (Sun MeanEcliptic J2000)

| Epoch (Sep. 23, 2025) | Before Impact | After Impact | Amount of Change |
|---|---|---|---|
| Semimajor axis, a (AU) | 0.922369289714061 | 0.922369674720639 | 3.9e-7 |
| Eccentricity, e | 0.191167354477951 | 0.1911659557505984 | -1.4e-6 |
| Inclination, i (°) | 3.340989755076949 | 3.340994018347669 | 4.3e-6 |
| RAAN, Ω (°) | 203.8997383804165 | 203.8997328405771 | -5.5e-6 |
| Argument of perihelion, ω (°) | 126.6764883710949 | 126.6765206259708 | 3.2e-5 |
| True anomaly, θ (°) | 228.9928359171914 | 228.9928091927391 | -2.6e-5 |

## Classic Kinetic Impactor Trajectory Design

For the CKI transfer model, the flight sequence is shown in **Figure 5**.

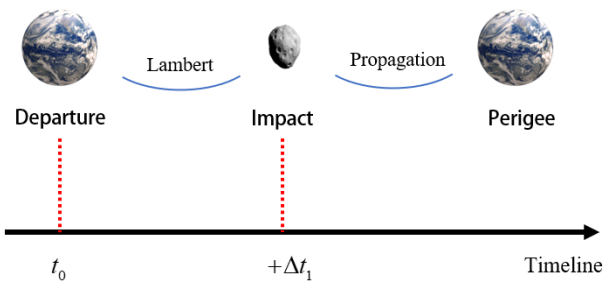

**Figure 5.** CKI transfer model. The spacecraft departs from Earth at $t_0$, impact the asteroid after $\Delta t_1$ days. By solving the Lambert problem, the hyperbolic excessive velocity $\boldsymbol{v}_\infty$ at Earth's SOI and the impact velocity $\boldsymbol{v}_{CKI}$ can be calculated. The $\boldsymbol{v}_\infty$ can be provided by launch vehicle, and the corresponding CKI mass can be calculated by launch capability.

This modeling process refers to Liu, et al.[43]. Assuming the CKI is launched directly into the impact orbit without any $\Delta v$ during the transfer trajectory, there are two decision variables in the optimization model $X = [t_0, \Delta t_1]$. To maximize the PHA's minimum geocentric distance during the close encounter, the objective function can be described as

$$J = \|\Delta \boldsymbol{r}'(t'_p)\| - \|\Delta \boldsymbol{r}(t_p)\| \tag{8}$$

where $\Delta \boldsymbol{r}'(t'_p)$ and $\Delta \boldsymbol{r}(t_p)$ indicates the PHA's perigee vector after and before the deflection. The genetic algorithm (GA)[44] is an evolutionary computational technique. Because of its global optimization ability, GA is used for the CKI mission optimization in this paper. **Figure 6**. shows the optimization process of a CKI mission. **Figure 7**. shows the CKI mission's transfer trajectory.

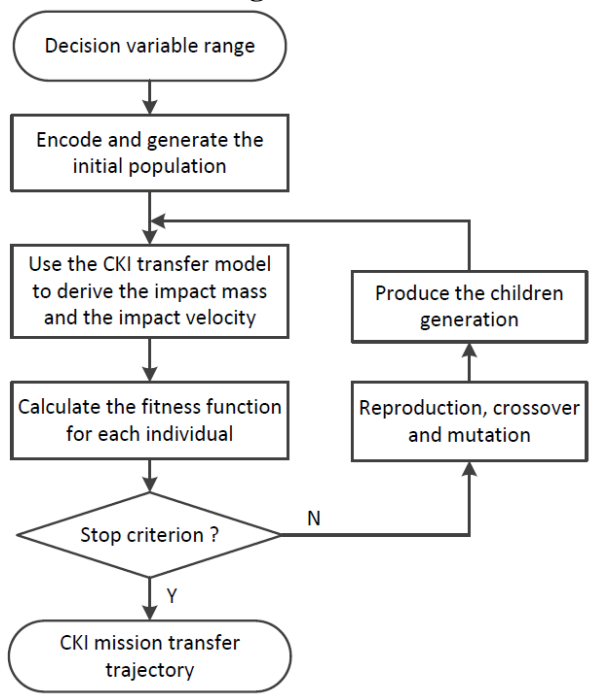

**Figure 6.** Optimization process of the CKI mission

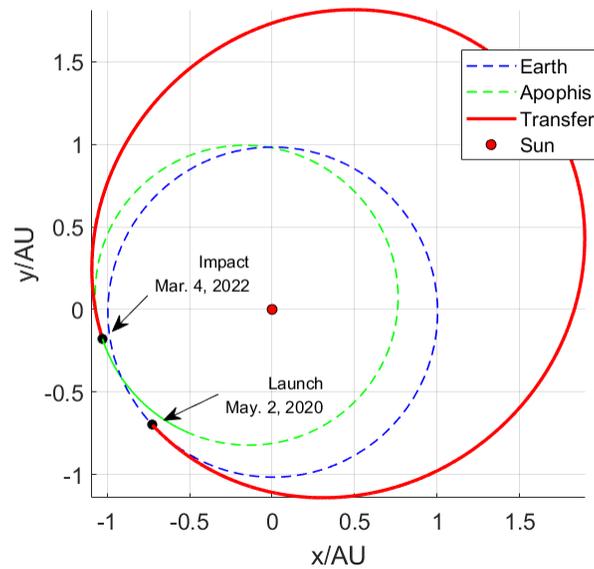

**Figure 7.** The CKI mission transfer trajectory. The red dot indicates Sun. Blue and green dotted lines indicate the orbits of Earth and Apophis. The red solid line indicates the CKI transfer trajectory. The actions of launch and impact are noted in this figure.

**Figure 1.** Schematic diagram of the Enhanced Kinetic Impactor (EKI). Four key actions are partially magnified in yellow circles: Launch, Rendezvous with NEA & Collect Rocks, Maneuver, and Impact PHA. The details of the NEA, PHA and spacecraft are magnified in gray circles. The gray and red dotted lines indicate the original orbits of the NEA and PHA, respectively. Solid lines of different colors represent transfer orbits at different stages.

**Figure 2.** Distance between Apophis and Earth. The blue dotted line indicates the geocentric distance before Apophis is impacted. The red solid line indicates the geocentric distance after Apophis is impacted. The black

dot on the left indicates the point at which Apophis is impacted by the EKI. The black dot on the right shows the details of the change in Apophis' minimum geocentric distance

**Figure 3.** Fuel-optimal low-thrust transfer trajectories. The red dot indicates Sun. Blue, gray and magenta dotted lines indicate the orbits of Earth, 2017 HF, and Apophis. Red solid lines indicate low-thrust trajectories. The four key actions mentioned in **Figure 1** are noted in the figure.

**Figure 4.** Propellant mass as a function of epoch. The two dotted lines divides the figure into three parts, which allows to visualize the propellant cost of the three arcs. The remaining propellant (1.02 t) can be used for other operations (de-spin et. al) and the mass of EKI.

**Figure 5．** CKI transfer model. The spacecraft departs from Earth at $t_0$, impact the asteroid after $\Delta t_1$ days. By solving the Lambert problem, the hyperbolic excessive velocity $\boldsymbol{v}_\infty$ at Earth's SOI and the impact velocity $\boldsymbol{v}_{CKI}$ can be calculated. The $\boldsymbol{v}_\infty$ can be provided by launch vehicle, and the corresponding CKI mass can be calculated by launch capability.

**Figure 6.** Optimization process of the CKI mission

**Figure 7.** The CKI mission transfer trajectory. The red dot indicates Sun. Blue and green dotted lines indicate the orbits of Earth and Apophis. The red solid line indicates the CKI transfer trajectory. The actions of launch and impact are noted in this figure.

## Acknowledgments

We thank Chi Wang, Xin Meng, Ji Wu, Zhen Yang et al. from National Space Science Center (NSSC, CAS) for their valuable discussions and suggestions; professor Jinguo Liu from SHENYANG Institute of Automation for discussing the cutting technology of rocks on asteroids; professor Qunbo Lv from Academy of Opto-Electronics for discussing the observation of small objects; Yamin Wang, Chengbo Qiu from Academy for Microsatellites of CAS for discussing the maneuver of small objects. We thank the simulation platform provided by Key Laboratory of Electronics and Information Technology for Space System, Chinese Academy of Science. We thank ARM's team for their valuable studies for technical issues of an asteroid capture mission. At the same time, we thank the reviewer and editor, who are working for Scientific Reports, for their reasonable and constructive suggestions.

## Author Contributions

Mingtao Li (corresponding author) and Yirui Wang proposed the concept of the Enhanced Kinetic Impactor (EKI). Mingtao Li guided the simulations and the article writing. Yirui Wang completed the simulations of the EKI mission and finished the article. Youliang Wang provided technical supports for low-thrust trajectory optimization. Binghong Zhou and Wei Zheng participated in the demonstration of the EKI.

## Additional Information

**Competing Interests:** The authors declare no competing interests